# Asteroids Detection Tehnique: Classic "Blink"
## An Automated Approch

*Denisa Copândean,*
*Constantin Nandra, Dorian Gorgan,*
Computer Science Department
Technical University of Cluj-Napoca
Cluj-Napoca, Romania
{denisa.copandean, constantin.nadra ,dorian.gorgan}
@cs.utcluj.ro

*Ovidiu Văduvescu*
Isaac Newton Group of Telescopes (ING),
Santa Cruz de la Palma, Canary Islands, Spain

Instituto de Astrofisica de Canarias (IAC),
La Laguna, Canary Islands, Spain
ovidiu.vaduvescu@gmail.com

*Abstract*—Asteroids detection is a very important research field that received increased attention in the last couple of decades. Some major surveys have their own dedicated people, equipment and detection applications, so they are discovering Near Earth Asteroids (NEAs) daily. The interest in asteroids is not limited to those major surveys, it is shared by amateurs and mini-surveys too. A couple of them are using the few existent software solutions, most of which are developed by amateurs. The rest obtain their results in a visual manner: they "blink" a sequence of reduced images of the same field, taken at a specific time interval, and they try to detect a real moving object in the resulting animation. Such a technique becomes harder with the increase in size of the CCD cameras. Aiming to replace manual detection, we propose an automated "blink" technique for asteroids detection.

## I. INTRODUCTION

Asteroids are not just a source of information about the origin of our Solar System or an opportunity for mining resources and modeling new frontiers for emerging space industry, they also represent a real threat for Earth, some of them having a high risk of collision with our planet. Near Earth objects (NEOs) discovery is a very difficult task because they have a small size, poor magnitude and some of them have move at high speeds. Monitoring the nearby space for near Earth asteroids is essential for the future of our planet. Planetary defense is the term used to encompass all the capabilities needed to detect the possibility of potential asteroid or comet impacts with Earth and to warn the population, its' purpose to either prevent these impacts or to mitigate their possible effects [1].

NASA (National Aeronautics and Space Administration) has been studying NEOs since the 1970s. The Agency initiated a survey, commonly called "Spaceguard," in the 1990s to begin searching for them. NASA participated in the International Spaceguard Survey, initiated in 1996 and sponsored by the multinational Spaceguard Foundation. NASA's NEO Observations Program supports NEO surveys that contribute to a sustained and productive campaign to find and track NEOs, collecting data of sufficient precision to allow accurate predictions of the future trajectories of the discovered objects. The Program also supports efforts to characterize a representative sample of NEOs by measuring their sizes, shapes, and compositions. NASA's Authorization Act of 2005 goal is to detect, track, catalogue, and characterize the physical characteristics of 90 percent of the NEO population with diameter sizes larger than 140 meters by 2020. While the major American surveys have increased the discovery rate, in Europe there are just a few local initiatives led by some scientists, without specific funding or dedicated facilities.

Recently, ESA (European Space Agency) has begun to contract some European services to contribute to SSA (Space Situational Awareness) program. NEODyS at the University of Pisa, KLENOT at the Klet Observatory, some national space agencies and other space industry entities are involved. Over the past few years, the SSA program used the one-meter telescope ESA-OGS in Tenerife for some NEO studies and further analysis [2][3]. The European Near Earth Asteroids Research (EURONEAR) project has been contributing to this research since 2006. EURONEAR includes only a small number of professional astronomers, involving more amateurs and students who promptly reduce the data, report discoveries and perform NEA recoveries. By manually "blinking" a set of consecutive images of the same field, they made 9 important discoveries and multiple of asteroids recoveries. Their actions

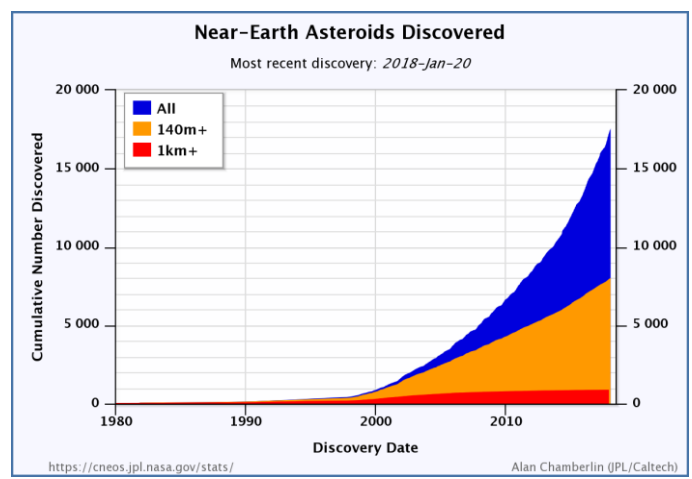

Fig. 1. NEA discoveries in time (https://cneos.jpl.nasa.gov/stats/totals.html)

were not part of an official NEA detection survey.

Having this manual flow in mind, we propose an automated pipeline prototype for moving object detection. The prototype is a modular system written in Python programming language, except for the Cross Object module which is written in Java programming language. Each module has a specific functionality, and some of them are tightly coupled to some other 3rd party astrophysics libraries.

The next section will presents some similar systems. Section 3 will consist of a short description of the correction modules for the proposed automated pipeline prototype and a more detailed description of the asteroids detection technique used in Cross Object module. Some results and the influence of the most important parameters involved in the detection prototype are presented in Section 4. The last section concludes on the achievements and experiments, presents the work currently in progress, and sketches some future research directions.

## II. RELATED WORKS

While making a star map, the Italian priest and astronomer Giuseppe Piazzi accidentally discovered the first asteroid, Ceres, orbiting between Mars and Jupiter. Today, Ceres is classified as a dwarf planet, having been considered the missing planet of the Bode's law. Bode's law suggested there should be a planet between Mars and Jupiter. It was discovered based on its motion relative to the fixed background stars in a couple of nights.

The same approach has been used from then on in asteroids detection, with evolved versions supported by the current time's technologies. In 1984, the Spacewatch group at the University of Arizona's Lunar and Planetary Laboratory has used this approach to implement the first computer detection system based on CCD (change coupled device) [5]. Their automated "Moving Object Detection Program" (MODP) searched for objects showing consistent motion in three successive scans over the same region. Each scan was 30 minutes long. During the first and second pass, their MODP recognized fast moving objects by the trails they left in the image. A moving object was detected on the third pass based on the displacement of its position in the first two passes relative to its position in the third pass [6]. So they developed a drift-scan technique.

Starting with the Near-Earth Asteroid Tracking (NEAT) program in 1995 a "step-stare" technique was implemented for asteroids detection [7]. A step-stare system takes multiple images, each being exposed while the telescope tracks along the Earth rotation. Then it was transferred to a computer before the next exposure was taken. This technique produced better results than a drift-scan technique and so it was further used by other surveys that followed. The Lincoln Laboratory's Near-Earth Asteroid Research (LINEAR) program run by the MIT Lincoln Laboratory takes for their detection software five images, 30 minutes apart. It only requires three hits for a detection [8].

Variable KD-Tree algorithms were used by the Pan-STARRS Moving Object Processing System [9], [10]. These techniques were developed in collaboration with the incoming LSST (Large Synoptic Survey Telescope) [11].

It was not only the major surveys that developed automated systems for moving object detection, but also some amateurs and small private surveys. For example, a group from Argentina developed such a system based on the profile of each light source represented by FWHM (Full Width at Half Maximum) [12].

## III. AUTOMATED PROTOTYPE FOR ASTEROID DETECTION

The current version of the proposed pipeline prototype is tightly coupled with the data obtained from the 2.5 meters diameter Isaac Newton Telescope (INT) located in La Palma, Canary Islands, Spain. At the prime focus of the INT, a wide field camera (WFC) is located, consisting of four CCDs of 2k x 4k pixels, covering an L-shaped 34 arcminutes x 34 arcminutes field with a pixel scale of 0.33 arcseconds pixel$^{-1}$.

### A. Short Introduction on the Pre-process Modules

Figure 2 presents the architecture of the prototype. The first two modules were developed in the Python programming language are based on a couple of external (3rd party) astrophysics libraries for image reduction and field correction [13].

Any image taken by a CCD camera through a telescope will not give accurate information about the photometric distribution over a portion of the sky. Optical imperfections and the discrete nature of light itself lead to errors in the measured data. The first module corrects artifacts (bias, flat

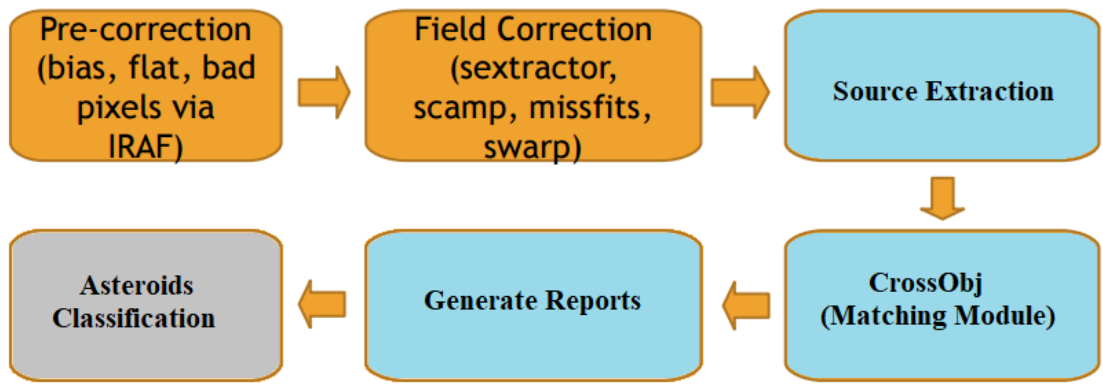

Fig. 2. Prototype Architecture

field) and instrumental defects (bad pixels) from the raw.

The second module of the prototype solves the field distortion issue of the reduced images in a sequence of steps based on a few external astrophysics libraries. Substantial field distortions caused by the optical system and projection of a hemispherical field into a 2D image occur in the imaging process, especially at the prime focus of any larger field telescope, which makes it difficult to perfectly align images for further processing and analysis. Field distortions increase with distance from the center of the image.

The first two blue-colored modules from the architecture described in figure 2, written in Java programming language represent the actual asteroid detection mechanism. Based on the findings, a report is generated under a specific format that is used by Minor Planet Center (MPC). Before submitting the report to MPC the list of known asteroids needs to be checked. In order to do that, the last module will query an external catalog for each detection and will adjust the report accordingly. This module will be soon implemented.

*B. General aspects used in CrossObj module*

Atmospheric turbulence contributes significantly to the degradation of the optical seeing quality. Astronomical seeing refers to the blurring and twinkling of astronomical objects, like stars, due to the turbulence in Earth's atmosphere, causing variations of the optical refractive index. As the light wave propagates through the atmosphere, it experiences fluctuations in amplitude and phase. There are several ways to quantify and evaluate seeing: Fried parameter, full width half maximum, Strehl intensity ratio, diameter enclosing with a specific percentage of the total energy and so on. Amateur astronomers evaluate seeing conditions using some astronomical seeing scales [14]. There are several types of scales. The measuring of angular sizes of the almost pointy stars is more professional than measuring by astronomical seeing scales. The angular diameter at half height of the star image profile is generally called the point spread function (PSF). Full Width Half Maximum (FWHM) is used to describe a measurement of the width of an object in a picture, when that object does not have sharp edges. The image of a star in an astronomical picture has a profile which is closer to a Gaussian curve. The width across the profile, when it drops to half its peak, or maximum value, is one of the best ways to reflect the approximate size of the star's image as seen by the human eye. FWHM serves in astrophotography as measurement of the astronomical seeing in arcseconds and it directly relates to the PSF of a seeing limited telescope. A bigger value of FWHM means worse conditions for observing. The FWHM is measured for a selection of stars in the frame and the "seeing" or image quality is reported as the mean value.

In our case the stars that are taken into account in order to compute the FWHM of the frame are those identify by SExtractor in the pre-process modules. The specific catalog of the frame that has the smallest FWHM value will be considered the reference catalog in the CrossObj module. Having the best astronomical seeing quality most of the asteroids should be available in that catalog.

The equatorial coordinate system is a celestial coordinate system used to specify the positions of celestial objects. The equatorial coordinate system is using spherical coordinates. Earth is at the center of the celestial sphere, an imaginary surface on which the planets, stars, and nebulae seem to be printed. The equatorial coordinate system is basically the projection of the latitude and longitude coordinate system we use here on Earth onto the celestial sphere. The right ascension (RA, symbol α) corresponds to east/west direction like longitude, while declination (DEC, symbol δ) measures north/south directions, like latitude. RA is measured in hours, minutes and seconds eastward along the celestial equator. The distance around the celestial equator is equal to 24 hours corresponding to 360 degrees (1h = 15 deg). DEC are measured in degrees, arcminutes and arcseconds, north or south of the celestial equator. There are 60 arcminutes in a degree, and 60 arcseconds in an arcminute and the DEC is running from -90 deg to +90 degree. Positive values for declination correspond to positions north of the equator, while negative values refer to positions south of the equator. The declination of the north celestial pole is 90° 0' 0" and the south celestial pole's declination is -90° 0' 0". The equator is 0° 0' 0".

RA and DEC are used in this approach in order to compute the imaginary distance between two objects on the celestial sphere.

$$d = arccos[sin\delta_1 * sin\delta_2 + cos\delta_1 * cos\delta_2 * cos(\alpha_1 - \alpha_2)] \quad (1)$$

where:

d - distance between two objects

$\delta_i$ - object declination

$\alpha_i$ - object right ascension

If the distance is less than 0.0004 units, those objects are considered to be the same object when it comes to background subtraction module. If the distance is less than 0.000025 units, two object are considered to be the same object when it comes to trajectory computation in CrossObj module. Distance between two objects is also used in proper motion calculation, needed in CrossObj module as a condition for pairing moving space objects.

Proper motion (symbol μ) refers to the angular velocity exhibited by a celestial body across the sky. It is the astronomical measure of the observed changes in the apparent places of stars in the sky, as seen from the center of mass of the Solar System, compared to the abstract background of the more distant stars [15]. The velocity of a star relative to the sun can be broken down into perpendicular components: the radial velocity and transverse velocity. The transverse velocity results in a change of angular position, which can be measured in arc seconds per year representing the proper motion of that star. So, proper motion may also be defined by the angular changes per year in the star's right ascension and declination, using a constant epoch in defining these. This is the approach taken in this prototype but with a time limit equal to the time frame between image acquisitions of the same observed field.

$$\mu = d / \Delta UT \quad (2)$$

ΔUT - time frame between image acquisitions of the same observed field

The proper motion has a magnitude and a direction. Its direction on the celestial sphere is given by position angle. The position angle (PA, symbol ω) is the convention for measuring angles on the sky in astronomy. The International Astronomical Union defines it as the angle measured counterclockwise, or direct sense, relative to the north celestial pole. As an angle it is measured in degrees: orientation to North is 0 deg, to East 90 deg, to South 180 deg, to West 270 deg. In the case of observed visual binary stars, it is defined as the angular offset of the secondary star from the primary relative to the north celestial pole. The definition of position angle is also applied to extended objects like galaxies, where it refers to the angle made by the major axis of the object with the north celestial pole line.

$$\theta = arctan(\Delta\delta, \Delta\alpha) \quad (3)$$

$$\omega = \theta + 2*\pi, \ \theta < 0 \quad (4)$$

$$\omega = \theta, \ \theta \geq 0 \quad (5)$$

Another characteristic of the space objects that is very useful in our prototype is the brightness. It represents the intensity of light received from a celestial object by an observer or apparatus. Because of the sensitivity of the human eye, brightness is perceived logarithmically, and the perceived intensity is measured in magnitudes. There are two different types: apparent magnitude, which is measured by an observer and absolute magnitude, which is taken at a standard distance and close to zero phase angle. Apparent magnitude (V - "visual") is observed and measured visually or with a CCD camera employing a suitable method to extract it [16]. Absolute magnitude (H) is the visual magnitude, V, an observer would record if the celestial object was placed 1 Astronomical Unit (AU) away, and 1 AU from the Sun and at a zero phase angle [17]. Very few asteroids have known sizes or shapes. Most of them have irregular shapes. Absolute magnitude H can be used to estimate the size of an asteroid if the albedo property is used too. Albedo is ratio of the light received by a body to the light reflected by that body. Albedo values range from 0, the pitch black, to 1, the perfect reflector. The brighter an object appears, the lower its magnitude value.

*C. CrossObj Module*

Background subtraction of celestial sources is an important phase for detecting moving objects. Stars and galaxies are considered fixed sources, fully populating the background field. If they can be eliminated, what remains are the moving objects (asteroids, comets and other Solar System objects) and the noise.

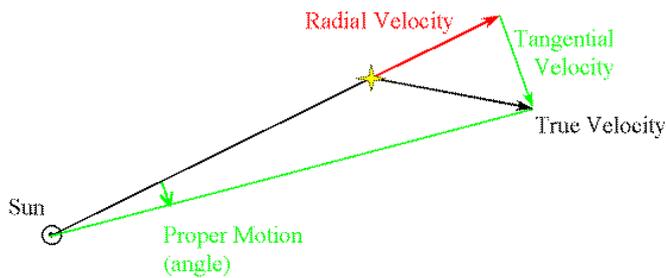

Fig. 3. Proper Motion (www.astronomy.ohio-state.edu)

The "Source Extraction" module loads the output file of the pre-process phases, the SExtractor ASCII catalog sources, as Java objects [13]. Those object will have all the properties discovered by SExtractor: FWHM, magnitude, RA, DEC and so on, needed further in the CrossObj module. This module will then iterate through all the catalogs specific to each field and each CCD and subtract the fixed objects that match sources from the combined catalog. The combined catalogs are obtained in the pre-process phases based on the median value of each pixel from the same position of the five frames. They contain only the fixed object. The subtraction is performed according to the distance computed with RA and DEC of the analyzed objects.

The matching module, CrossObj, is based on the classic *blink* algorithm. Before starting to pair the remaining celestial objects after background subtraction, the catalogs are sorted after the FWHM property. This parameter associated with a catalog can be interpreted as the clarity of the sky for the image that the catalog represents. The catalog of the image that have the best FWHM will be considered the pivot in the analyzed sequence of images (5 images). After that, for each source in the pivot, the algorithm will make pairs with sources from the others catalogs (pair coupling a source from the pivot with a source from another catalog). In order for two sources to be paired, they need to be at a certain distance in a known time interval (1 < d < 10 "/min) with a certain velocity (less than 10).

The actually matching is done in the second part of the module. If four pairs having the same source from the pivot comply with both the direction (based on the position angle) and proper motion, then it is safe to assume that the object describing such a trajectory is an asteroid. Equation 6 represents the mathematical condition for two pairs to describe a trajectory. This condition must be valid for all combinations of the four pairs that define a detection. More than that, all space objects from the pairs must be from different catalog, except for the pivot source that will appear in all four pairs.

$$|\omega_{pair1} - \omega_{pair2}| < 10^o \text{ and } |\mu_{pair1} - \mu_{pair2}| < 1''/min \quad (6)$$

All NEO search and tracking projects are required to make their data permanently available in a timely manner to the scientific community. Observers provide their data to a global database maintained by the Minor Planet Center, which is internationally recognized as the public archive for these data. MPC is sanctioned by the International Astronomical Union and supported by the NEO Observations Program [1]. The Minor Planet Center is tasked with notifying observers worldwide about PHOs so they can conduct timely follow-up observations.

So, from the outcome of the detection module, CrossObj, a MPC report is automatically generated. This report has a strict format and it must contain a generic identifier for each detection, the observation date, the source location on the sky (RA and DEC) and its magnitude along with the character "R" and the observer identifier at the end of line. Each source that defines a detection is described in a chronological way, per asteroid.

## IV. Observed Influence of the Most Important Parameters

The first round of tests was done against a set of archives provided by the collaborative project EURONEAR. In this collection, the time frame between image acquisitions of the same observed field was very small, and some of the asteroids identified by the human eye were flagged as part of the background in the prototype. So the time acquisition is very important in classifying a source as static or moving. For the second round of tests, the time between acquisitions of the same field was increased, letting the asteroids to move across the sky.

The astronomical seeing, the FWHM property, of the individual space object was used to filter the sources that are loaded in Java memory for each catalog, except for the combined catalogs. An object that has the FWHM under 0.5 is a noise. After this condition was updated from 0 to 0.5, the process's time decreased significant from tree-four minutes to just one minute. More than that, the number of false identified trajectories was reduced from thousands to less than 20.

But not only FWHM produced such an improvement. It was a mixt between it and a condition over the magnitude too. Initially, it was set to eliminate saturated stars, but then it became apparent that it could also filter noise by allowing only sources with magnitude in interval 0 and 30. With these settings 16 detections, with 9 positives and 7 false matches, were obtained. Studying the nature of the asteroids it was decided to restrict the magnitude limits, first to [15, 26], when 10 asteroids out of 12 identified trajectories were discovered, then to [17, 24], when along with another adjustment of FWHM to 0.7 a score of 11 out of 11 was obtained.

Usually, in a field, there should be 30-40 asteroids. Experiments need to be conducted in order to determine the influence of the parameters in the pre-process phases as well, in order to access sources that right now are not accessible.

## V. Conclusions and Future Directions

Medium and large sized telescopes (2-4 meters) are required to detect faint NEAs using the classic "blink" algorithm and in the near future it is planned to use this pipeline for such data. Even so, smaller telescopes (1 meter) can be used in the same purpose if the used process is based on various forms of "digital tracking" methods that require extensive computing resources. This is another direction that the prototype will approach in the future.

Having a modular structure, the prototype can be easily improved from the performance point of view. It will be migrated to cloud for high performance computing. A cloud solution based on the Docker and Kubernetes technologies is currently under development. This will eliminate all the problems encountered in the setup process of the environment. It will also bring a great value to any user that will want to use this prototype.